\documentstyle[11pt]{article}

\begin{document}
\thispagestyle{empty}

\begin{center}
%%%%%%%%%%%%%%%%%%%%%%%%%%%%%%%%%%%%%%%%%%%%%%%%%%%%%%%%%%%%
{\large \bf Phase fluctuations and Non-Fermi Liquid  Properties of\\
2D Fermi-system with attraction}
%%%%%%%%%%%%%%%%%%%%%%%%%%%%%%%%%%%%%%%%%%%%%%%%%%%%%%%%%%%%
\end{center}

\begin{center}
Valery~P.~Gusynin$^{1}$,
        Vadim~M.~Loktev$^{1}$,
                     Rachel~M.~Quick$^{2}$
 and \underline{Sergei~G.~Sharapov}$^{2, \dagger}$\\
%%%%%%%%%%%%%%%%%%%%%%%%%%%%%%%%%%%%%%%%%%%%%%%%%%%%%%%%%%%%
$^{1}${\sl Bogolyubov Institute for
               Theoretical Physics,}\\
         {\sl  252143 Kiev, Ukraine} \\
$^{2}${\sl Department of Physics,
               University of Pretoria,}\\
            {\sl 0002 Pretoria, South Africa}
\end{center}

\noindent
%%%%%%%%%%%%%%%%%%%%%%%%%  ABSTRACT   %%%%%%%%%%%%%%%%%%%%%%%%%%%%%%%
The effect of static fluctuations in the phase of the
order parameter on the normal and superconducting properties of
a 2D system with attractive four-fermion interaction has been
studied. Analytic expressions for the fermion Green function, its
spectral density and the density of states are derived. The
resultant single-particle Green function clearly demonstrates
non-Fermi liquid behavior. The results show that as the temperature
increases through the 2D critical temperature the width of the
quasiparticle peaks broadens significantly. At the same time one
retains the gap in quasiparticle spectrum. The spectral density for
the dynamical fluctuations can also be obtained. Clearly the dynamical
fluctuations fill the gap giving the observed pseudogap behaviour.\\
%%%%%%%%%%%%%%%%%%%%%%%%%%%%%%%%%%%%%%%%%%%%%%%%%%%%%%%%%%%%%%%%%%%%%

%%%%%%%%%%%%%%%%%%%%%%%%%%%%%%%%%%%%%%%%%%%%%%%%%%%%%%%%%%%%
% PACS 74.72.-h, 74.20.Fg, 74.20.Mn, 64.60.Cn
%%%%%%%%%%%%%%%%%%%%%%%%%%%%%%%%%%%%%%%%%%%%%%%%%%%%%%%%%%%%

\noindent
$^{\dagger}$
On leave from Bogolyubov Institute for Theoretical Physics
of the National Academy of Sciences of Ukraine, 252143 Kiev,
Ukraine \\

\noindent
Invited paper presented at the Second International Conference on\\
{\em New Theories, Discoveries, and Applications of Superconductors
and Related Materials}\\
May 31-- June 4, 1999, Las Vegas, Nevada, USA

\vspace{2cm}

%\newpage
It is known \cite{Emery} that the phase fluctuations are especially
important in low dimensional superconductors with small carrier
density (such as the high-$T_c$ oxide superconductors). In particular
there is some experimental evidence that the transition into
superconducting state in high-$T_c$ cuprates is driven by
phase fluctuations.\cite{Corson} The microscopic theory developing
this scenario \cite{Emery} has been proposed in \cite{we}.

The main quantity of interest in the present report is the
one-fermion Green's function and the associated spectral function
$A(\omega,{\bf k})=-(1/\pi){\rm Im}G(\omega+i0,{\bf k})$. The
second quantity, being proportional to the intensity of the
angle-resolved photoemission spectrum (ARPES) \cite{Campuzano},
encodes information about the quasiparticles.
Following the approach of Ref.\cite{we} the Green's function for
the charged (physical) fermions, $G(\omega, {\bf k})$ is given by
the convolution (in momentum space) of the propagator for neutral
fermions which has a gap $\rho\neq0$ and the Fourier transform of
the phase correlator
$\langle\exp(i \theta(x)/2)\exp(-i\theta(0)/2)\rangle$.

We demonstrate that the quasiparticle spectral function broadens
considerably when passing from the superconducting to the normal
state as observed experimentally \cite{Campuzano}. More importantly
the phase fluctuations result in non-Fermi liquid behavior of the
system both below and above $T_{\rm BKT}$, where $T_{\rm BKT}$
is the temperature of Berezinskii-Kosterlitz-Thouless (BKT)
superconducting transition in 2D.

In the frequency-momentum representation the Green function of
charged fermions can be written (see details in \cite{we.new})
as
\begin{equation}
G(i \omega_{n}, \mbox{\bf k}) = T \sum_{m = - \infty}^{\infty} \int
\frac{d^2 p}{(2 \pi)^{2}}
\sum_{\alpha,\beta=\pm}
P_\alpha {\cal G}(i \omega_{m}, \mbox{\bf p}) P_\beta D_{\alpha
\beta} (i \omega_{n} - i \omega_{m}, \mbox{\bf k} - \mbox{\bf p}),
\label{Green.projectors.momentum}
\end{equation}
where ${\cal G}(i \omega_{m}, \mbox{\bf k})$ is the Green function
of neutral fermions;
$P_{\pm} \equiv (1/2) (\hat I \pm \tau_3)$ are the projectors;
$\hat I$, $\tau_3$ are unit and Pauli matrices;
$D_{\alpha \beta}(i \Omega_{n}, \mbox{\bf q})$
is the Fourier transform of the correlator of the phase fluctuations
$\langle \exp [ i\alpha \theta(x) / 2 ]
\exp [ - i\beta \theta(0) / 2 ] \rangle$ ($x = \tau, \mbox{\bf r}$
with the imaginary time $\tau$ and 2D space coordinate $\mbox{\bf r}$);
$\omega_n = (2n+1) \pi T$, $\Omega_{n} = 2 \pi n T$ are fermion (odd)
and boson (even) Matsubara frequencies.

One can show that $D_{+-} = D_{-+} = 0$ which retains the symmetric
form of the propagator $G$ is accordance to the
Coleman-Mermin-Wagner-Hohenberg theorem and the properties of
(\ref{Green.projectors.momentum}) are entirely determined by
$D \equiv D_{++} = D_{--}$.

The general form of the phase
correlator is
\begin{equation}
D(t, \mbox{\bf r}) = \exp (- \gamma t)
\left( \frac{r}{r_{0}} \right)^
{-\frac{\displaystyle T}{\displaystyle 8 \pi J}}
\exp \left( - \frac{r}{\xi_{+}(T)} \right),
\label{correlator.dynamic}
\end{equation}
where $\gamma$ is the phenomenological decay constant;
$J(T)$ is the superfluid stiffness \cite{we};
$r_0$ is the scale of algebraic decay of the correlations in the
BKT phase ($T < T_{\rm BKT}$); $\xi_{+}(T)$ is the correlation
length for $T > T_{\rm BKT}$ ($\xi_{+} = \infty$ for
$T \leq T_{\rm BKT}$).
Note that $t$ is the real time, so that (\ref{correlator.dynamic})
is the retarded Green function. The power decay factor in Eq.
(\ref{correlator.dynamic}) is related to the spin-wave
phase fluctuations, while the exponent with $\xi_{+}$ is responsible
for the vortex excitations, which are present only above
$T_{\rm BKT}$.

It is remarkable that the static case ($\Omega_n = \gamma = 0$) can be
studied analytically and  one obtains the retarded
Green function \cite{we.new}
\begin{eqnarray}
G(\omega,\mbox {\bf k}) =
- \frac{C m \xi_{+}^{2\alpha}}{2\pi\alpha}
&& \! \! \! \! \! \! \! \!
\left[\frac{\cal A}{(v_1 v_2)^{\alpha}} F_1 \left(\alpha, \alpha, \alpha;
\alpha + 1; \frac{v_1 - 1}{v_1}, \frac{v_2 - 1}{v_2} \right)
\right. \nonumber         \\
&& \left.
+ (\sqrt{\omega^{2} - \rho^{2}} \to -\sqrt{\omega^{2} - \rho^{2}})
\right],
\label{Green.final}
\end{eqnarray}
where $F_1$ is the Appell's function of two variables,
\begin{eqnarray}
&& \alpha =  1 - \frac{T}{16 \pi J(T)}, \quad
C = \frac{4 \pi \Gamma(\alpha)}{\Gamma(1-\alpha)}
\left( \frac{2}{r_0} \right)^{2\alpha -2}, \quad
{\cal A} = \frac{1}{2} \left( \tau_3  +
\frac{\omega}{\sqrt{\omega^2 - \rho^2}} \right),
\nonumber        \\
&& v_{1,2} = m \xi_{+}^{2}(v_0 \pm \sqrt{\cal D}), \quad
{\cal D} = v_{0}^{2} +  \frac{2}{m \xi_{+}^{2}}
(\mu + \sqrt{\omega^{2} - \rho^{2}}),
\nonumber                       \\
&& v_0 = \frac{k^2 \xi_{+}^{2} + 1}{2m \xi_{+}^{2}} - \mu
-  \sqrt{\omega^{2} - \rho^{2}},
\label{alpha}
\end{eqnarray}
$m$ is the mass of fermions.

For $T < T_{\rm BKT}$ near the quasiparticle peaks when
$\omega \approx \pm \sqrt{(\mbox{\bf k}^2/2m - \mu)^{2} + \rho^2}$
one can derive from (\ref{Green.final}) that
\begin{equation}
G(\omega, \mbox{\bf k})  \sim -\Gamma^{2}(\alpha)
\left(\frac{2}{m r_{0}^{2}}\right)^{\alpha - 1}
\frac{{\cal A}}{[- (\mu + \sqrt{\omega^{2} - \rho^{2}})]^\alpha}
\frac{\Gamma(2\alpha-1)}{\Gamma^{2}(\alpha)}
\frac{1}{(1-z_1)^{2\alpha - 1}},
\label{cut.<}
\end{equation}
where $z_{1} \equiv (k^2/2m)/ (\mu + \sqrt{\omega^{2} - \rho^{2}})
\simeq 1$. It is seen that Eq. (\ref{cut.<}) is evidently
nonstandard since it  contains a branch cut, not pole.
The latter in its turn corresponds to the non-Fermi liquid behavior
of the system as whole. It must be underlined that non-Fermi
liquid peculiarities are tightly related to the charge
(i.e. observable) fermions only -- the Green function of neutral
ones has typical (pole type) BCS form.  Besides, because the parameter
$\alpha$ is a function of $T$ (see (\ref{alpha})) the non-Fermi
liquid behavior is becoming more pronounced as temperature increases
and is preserved until $\rho$ vanishes.

The spectral function $A(\omega,{\bf k})$ associated with
$G(\omega,{\bf k})$ has been studied in \cite{we.new}.
Here we sketch the major features.
$i)$ The quasiparticle peaks of the spectral density have
a finite temperature dependent width which decreases with
decreasing $T$ in the superconducting state. The
sharpening of the peaks with decreasing $T$ is seen
experimentally \cite{Campuzano} and this is in a striking
difference from the BCS temperature independent ``pile-up''.
$ii)$ For $T > T_{\rm BKT}$ the width of peaks broadens
dramatically which is also observed experimentally \cite{Campuzano}.
$iii)$ In the static approximation ($\Omega_n = \gamma=0$) we
obtained two extra peaks which are present for $T \neq 0$. The
gap in the excitation spectrum remains nonfilled even above
$T_{\rm BKT}$.
$iv)$ The spectral density for the dynamical fluctuations, i.e.
when $\Omega_{n} \neq 0$ modes are taken into account
can also be obtained.
The extra peaks seen in the static case disappear.
Clearly the dynamical fluctuations fill the gap giving the
observed pseudogap behaviour \cite{Campuzano} where the
strenght of the filling depends on $\gamma$.

\section*{Acknowledgements}
\noindent
R.M.Q and S.G.Sh acknowledge the financial support of the
NRF, Pretoria.

\end{document}